\documentclass[prl,aps,amsmath,amssymb,twocolumn]{revtex4}
\usepackage{graphicx}
\usepackage{bm}
\begin{document}

\title{Two-impurity scattering  in quasi-one-dimensional systems}
\author{A. S. Ioselevich\footnote{e-mail: iossel@itp.ac.ru}}
\affiliation{Condensed-matter physics laboratory, National Research University Higher School of Economics, Moscow 101000, Russia,}
\affiliation{L. D. Landau Institute for Theoretical Physics, Moscow 119334, Russia}
\author{N. S. Peshcherenko\footnote{e-mail: peshcherenko@itp.ac.ru}}
\affiliation{L. D. Landau Institute for Theoretical Physics, Moscow 119334, Russia}

\address{}
\date{\today}

\begin{abstract}
In a quasi-one-dimensional system (a tube) with low concentration of  defects $n$ the resistivity $\rho$ has peaks (van-Hove singularities) as a function of Fermi-energy. We show that due to non-Born scattering effects a deep narrow gap  should appear just in the center of each peak. The resistivity at the bottom of a gap ($\rho_{\min}\propto n^2$) is dominated by scattering at rare ``twin'' pairs of close defects, while scattering at solitary defects is suppressed. The predicted effect is characteristic for multi-channel systems, it can not be observed in strictly one-dimensional one.
 \end{abstract}

\maketitle

In quasi-one-dimensional systems 
with low concentration of impurities the quantization of transverse electronic motion is essential, and the conductivity shows Van Hove singularities when the Fermi level approaches a bottom of some transverse subband \cite{VanHove}. A variety of such systems includes carbon nanotubes \cite{single-wall0,single-wall, multi-wall1,multi-wall2}, thin wires \cite{Brandt77,Nikolaeva2008}, nanoribbons \cite{nanoribbon1,nanoribbon2}  or long constrictions in 2D semiconductor heterostructures \cite{constrictions,Thornton1986,Zheng1986,stripstudyexper}.  In experiment the observed Van Hove singularities may have  quite complex
structure, which is often attributed to Fano resonances \cite{fano, fano1, single-wall} (see, however, a discussion in \cite{IosPeshJETPL2018}).

In our previous
 work \cite{IosPeshJETPL2018, IosPeshPRB2019, IosPeshPRB2020} we have demonstrated that in the central part of each singularity the scattering is  suppressed due to non-Born effects, so that the shape of singularity is dramatically modified. Qualitatively, the non-Born screening of scattering originates in destructive interference of multiply scattered waves with different winding numbers. 
  These results were restricted to the ``single-impurity approximation'' when the interference of scattering events at different impurities could be neglected. This  approximation, however,  breaks down in the immediate vicinity of the Van Hove singularity where multi-impurity effects become crucial and an essentially quantum approach is necessary. 
 In this paper we develop such an approach and show that the multi-impurity effects are effectively reduced to two-impurity ones. Moreover, the leading contribution to resistivity comes from anomalously close pairs of impurities. 
 
 For simplicity in this letter we consider only point-like repulsing impurities and only the case when impurities are effectively equivalent (it is so for a tube, but not for a strip, see \cite{IosPeshPRB2020}). Modifications arising in the case of attraction or in the case of strip will be discussed in an extended publication. 

\begin{figure}[ht]
\includegraphics[width=0.7\linewidth]{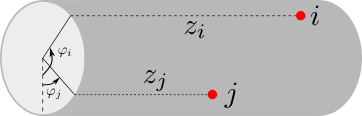}
\caption{A tube with two point-like impurities on its surface.}
\label{draft-tube}
\end{figure}

{\bf The model.} Following \cite{IosPeshPRB2019} we consider noninteracting electrons with  spectrum $E=\hbar^2k^2/2m^*$ on a tube of radius $R$ with identical point-like repulsing impurities randomly placed on its surface.  All distances below are measured in units of $2\pi R$, all energies -- in units of $\hbar^2/2m^*R^2$.  Hamiltonian of electrons in these units  is
\begin{eqnarray}
\hat{H}=-\frac{1}{(2\pi)^2}\frac{d^2}{dz^2}-\frac{\partial^2}{\partial\phi^2}+\frac{\lambda}{\pi^2}\sum_j\delta(z-z_j)\delta(\phi-\phi_j),
\label{Schrodinger_eq0}
\end{eqnarray}
where position of an electron is characterized by coordinates $z,\phi$, position of $i$-th impurity -- by randomly distributed uncorrelated $z_i,\phi_i$ (see Fig. \ref{draft-tube}).

The one-dimensional density $n$ (average number of impurities per unit length of a cylinder) is small $n\ll 1$. The dimensionless Born scattering amplitude $\lambda>0$ is also small ($\lambda\ll 1$). The spectra of transverse quantization subbands are $E_m(k)=m^2+k^2/(2\pi)^2$, where $m=0,\pm 1,\pm 2\ldots$ and $k$ is the one-dimensional momentum along the tube. A convenient method to control the position of the Fermi level is to thread the tube by a magnetic flux $\Phi$. Then
\begin{align}
E_m(k)=(m+\Phi/2\Phi_0)^2+k^2/(2\pi)^2,
\label{flux}
\end{align}
where $\Phi_0=\pi c\hbar/e$ is the flux quantum. Note that the presence of flux $\Phi\neq 2M\Phi_0$ somewhat simplifies our problem, lifting the initial degeneracy $m\to-m$.
The Fermi level $E_F$ is supposed to be close to the bottom of $N$-th subband: $E_F=(N+\Phi/2\Phi_0)^2+\varepsilon$, $|\varepsilon|\ll 1$, where the number $N$ of open channels $2N$ is large ($2N\gg 1$). The states in the $N$-th band have very low longitudinal velocity $\sim \sqrt{\varepsilon}$ and do not directly contribute to the current. Nevertheless, due to their high density of states, they play important role in resistivity as final (or intermediate) states in the processes of  scattering for current-carrying states from other bands.

We restrict our consideration to the tubes of intermediate lengths $l\ll L_{\rm tube}\ll\l_{\rm loc}$, where $l$ -- mean free path and $\l_{\rm loc}\sim Nl\gg l$ the localization length. This allows not to take into account weak localization corrections.

{\bf Previous results.} There are two  energy scales in this problem
\begin{align}
\varepsilon_{\rm nB}=\left(\lambda/\pi\right)^2,\quad \varepsilon_{\min}=\left(n/\pi\right)^2.
\label{scales-gen}
\end{align}
In the most interesting case of low concentration $\varepsilon_{\min}\ll \varepsilon_{\rm nB}$ the single-impurity non-Born effects  lead to effective screening of the renormalized scattering amplitude in the range of energies $\varepsilon_{\min}\ll\varepsilon\ll \varepsilon_{\rm nB}$. In \cite{IosPeshPRB2019} we have shown that
\begin{align}
\frac{\rho}{\rho_0}=\left\{\begin{aligned}\frac{1}{\lambda}\left[\left(\varepsilon/\varepsilon_{\rm nB}\right)^{1/2}+\left(\varepsilon_{\rm nB}/\varepsilon\right)^{1/2}\right]^{-1},&\quad\varepsilon>0,\\
\left[1+\left(\varepsilon_{\rm nB}/|\varepsilon|\right)^{1/2}\right]^{-2},&\quad\varepsilon<0,
\end{aligned}\right.
\label{old-results}
\end{align}
where $\rho_0=(4\pi/N^2e^2)n\left(\lambda/\pi\right)^{2}$ is the resistivity away from the Van Hove singularities. Thus, for $|\varepsilon|\ll \varepsilon_{\rm nB}$ the resistivity is suppressed ($\rho\propto\sqrt{\varepsilon}$), while for $|\varepsilon|\gg \varepsilon_{\rm nB}$ it is described by standard Born result ($\rho\propto1/\sqrt{\varepsilon}$).

{\bf New results.} The single-impurity approximation is applicable and the formula \eqref{old-results} is valid only for $|\varepsilon|\gg\varepsilon_{\min}$. In \cite{IosPeshPRB2019} we have estimated the resistivity in the range $|\varepsilon|\lesssim\varepsilon_{\min}$ with the help of ``self-consistent non-Born approximation'' that could not be reliably justified. In this letter we present a full analytical solution for the entire range $|\varepsilon|\ll\varepsilon_{\rm nB}$ (including $|\varepsilon|\lesssim\varepsilon_{\min}$). This solution is based on certain ideas and exact results from the theory of strictly one dimensional systems (see \cite{LifshitsGredeskulPastur1982,BychkovDykhne1966,BychkovDykhne1967}). These results, however, had to be considerably modified to reflect the multi-channel character of our problem. 

\begin{figure}[ht]
\includegraphics[width=0.9\linewidth]{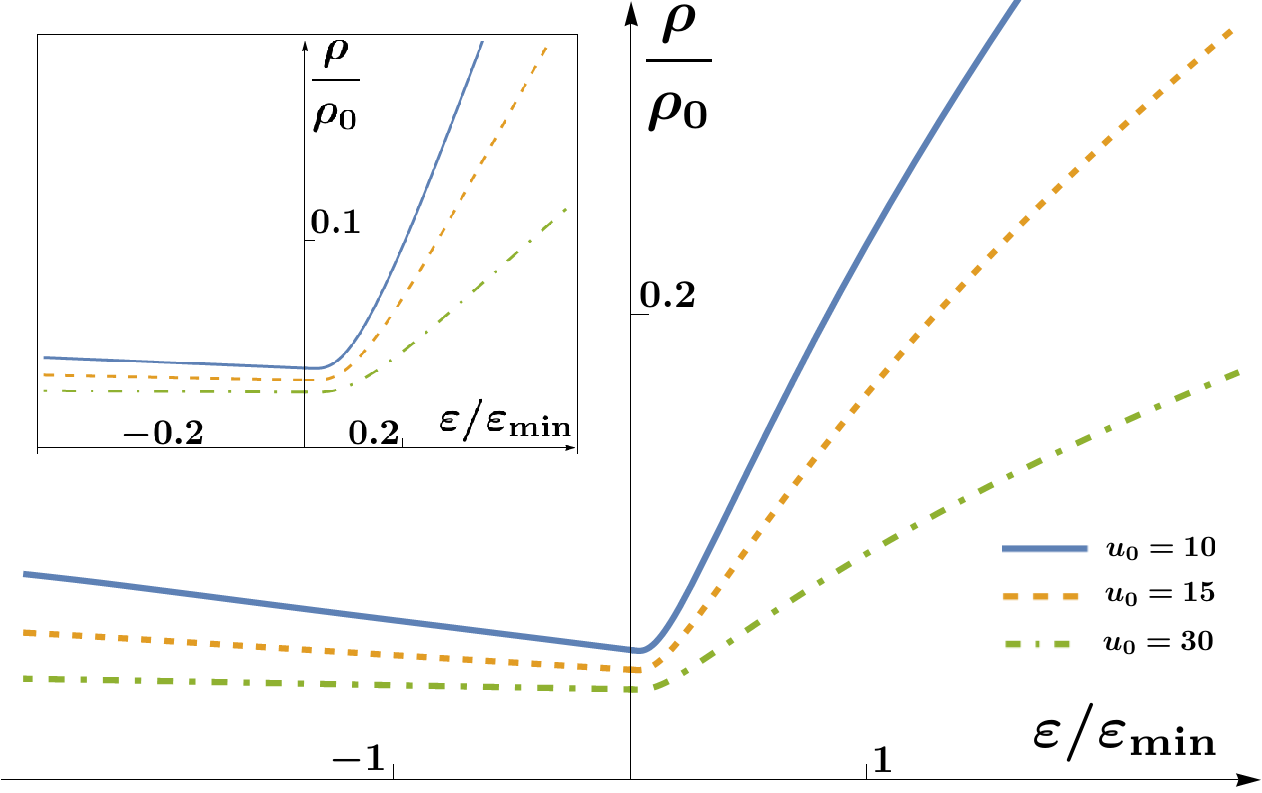}
\caption{Plot of the total resistivity $\rho(\varepsilon)$ for $\lambda=0.2$ (main plot) and $\lambda=0.05$ (inset). In both cases three values of $u_0=\lambda/n$ are used: $u_0=10,15,30$.
 }
\label{draft-resistivity}
\end{figure}

The main results of this paper are as follows. The resistivity consists of four parts:
$\rho=\rho_{\rm res}+\rho_{\rm twin}+\rho_{\rm typ}+\Delta\rho_{\rm log},$
\begin{align}
\frac{\rho_{\rm res}}{\rho_0}=\frac{\pi\theta(\varepsilon)/2\lambda u_0}{\exp\left(\pi/2u\right)-1}\approx\frac{\pi\theta(\varepsilon)}{2\lambda u_0}\left\{\begin{aligned}2u/\pi,\quad& u\gg 1,\\
e^{-\pi/2u},\quad& u\ll 1,
\end{aligned}
\right. \label{resuu01}
\\ \rho_{\rm twin}/\rho_0=1/4u_0,
\label{twinies}\\ 
    \frac{\rho_{\rm typ}}{\rho_0}=\frac{(-\varepsilon)}{\varepsilon_{\rm nB}}\left\{\begin{aligned}1,\quad& u\gg 1,\\
8/3,\quad& u\ll 1,
\end{aligned}
\right.
\label{resuu02}\\
\frac{\Delta\rho_{\rm log}}{\rho_0}=\frac{(-\varepsilon)}{\varepsilon_{\rm nB}}\left\{\begin{aligned}(1/2u)\ln (u_0/u)\quad& u\gg 1,\\
(1/3)\ln (u_0)\quad& u\ll 1,
\end{aligned}
\right.+\nonumber\\+\frac{1}{8u_0^2}\left\{\begin{aligned}\ln^2 (u_0/u)\quad& u\gg 1,\\
\ln^2 (u_0),\quad& u\ll 1,
\end{aligned}
\right.
\label{resuu03}
\end{align}
where we have introduced
\begin{align}
u=(|\varepsilon|/\varepsilon_{\min})^{1/2}\ll u_0,\quad  u_0=(\varepsilon_{\rm nB}/\varepsilon_{\min})^{1/2}\gg 1\label{resuu2}.
\end{align}
The total resistivity $\rho(\varepsilon)$ 
 is plotted in Fig. \ref{draft-resistivity}.

\begin{figure}[ht]
\includegraphics[width=0.8\linewidth]{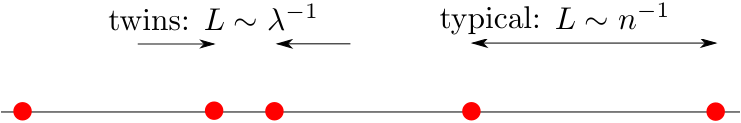}
\caption{Illustration: twin-pair versus typical pair.}
\label{Figure_3}
\end{figure}

1.  The part $\rho_{\rm res}$ is dominant in the range $\varepsilon_{\min}\ln^{-2}(1/\lambda)<\varepsilon\ll \varepsilon_{\rm nB}$. It is due to resonant scattering at pairs with such separations $L$ that one of the quasistationary energy levels $E_p\approx p^2/4L^2$ ($p=1,2,3,\ldots$) for an electron trapped between these two impurities is tuned to the energy $\varepsilon$ (see Fig. \ref{Figure_3}). If $\varepsilon\sim\varepsilon_{\min}$, (or $u\sim 1$) then relevant resonant pairs correspond to  typical $L\sim n^{-1}$ and $p\gtrsim 1$, 
while at $u\gg 1$ the relevant $p\gg 1$. For $1\ll u\ll u_0$  the result \eqref{resuu01} overlaps with the single-impurity one \eqref{old-results}. For $u\ll 1$ \eqref{resuu01} predicts an even stronger suppression of $\rho_{\rm res}$ than in \eqref{old-results}:   the relevant pairs with $L\sim\varepsilon^{-1/2}\gg n^{-1}$ are exponentially rare and therefore $\rho_{\rm res}$ is exponentially small.

 \begin{figure}[ht]
\includegraphics[width=0.8\linewidth]{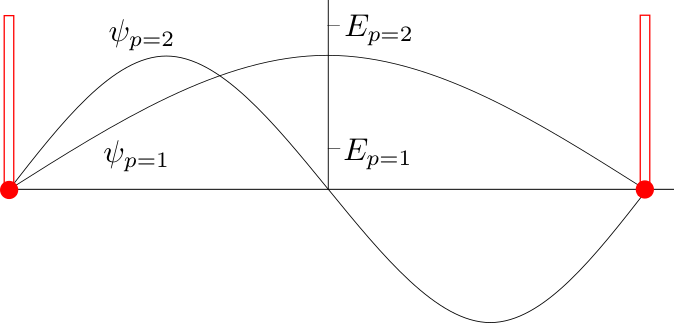}
\caption{Wave-functions $\psi_p$ and resonant energies $E_p$ for quasistationary states associated with a pair of impurities (their strongly repulsing potentials are shown as vertical bars).}
\label{Figure_2}
\end{figure}

2. The part   $\rho_{\rm twin}$ dominates in the range of smallest energies
\begin{align}
-(\varepsilon_{\min}\varepsilon_{\rm nB})^{1/2}<\varepsilon<\varepsilon_{\min}\ln^{-2}(1/\lambda).
\label{interval}
\end{align}
 It is the contribution of ``twins'' -- rare pairs of impurities with non-typically small spatial separation between components: $L\sim\lambda^{-1}\ll n^{-1}$ (see Fig. \ref{Figure_2}). 
The  twins are more  effective scatterers than solitary impurities because the non-Born screening effect  that perfectly suppresses the single-impurity scattering at $\varepsilon\to0$ (see \cite{IosPeshPRB2019, IosPeshPRB2020}), gradually becomes weaker as impurities come closer to each other, and almost vanishes  for twins. It is important that this phenomenon exists only in a multi-channel system ($N\gg1$). We note that somewhat similar (though with a completely different physical background) effect  -- the dominance of skew scattering at close pairs of impurities -- was also recently predicted for a 2D ferromagnet with strong spin-orbit interaction \cite{Ado2016, Ado2020}. Some results for the resistance of a strip with only two impurities in it were obtained in
\cite{KumBag1991}.

 3. The part $\rho_{\rm typ}$ dominates in the range of large negative energies $-\varepsilon_{\rm nB}\ll \varepsilon\ll-(\varepsilon_{\min}\varepsilon_{\rm nB})^{1/2}$. It is due to   nonresonant scattering at  typical pairs with $L\sim 1/n$. The corresponding result \eqref{resuu02} does not differ from the quasiclassical one \eqref{old-results}.
 
 4. The  $\varepsilon$-dependent term $\Delta\rho_{\rm log}$ at all energies is only a small correction to the other terms. Nevertheless, it is relevant since it is the principal source of the energy dependence of the resistivity in the range $-\varepsilon_{\min}\ln^2 u_0\ll\varepsilon<\varepsilon_{\min}\ln^{-2}(1/\lambda)$. This logarithmic contribution comes from a wide interval of distances between twins and typical pairs: $1/\lambda<L<\min\{1/n,|\varepsilon|^{-1/2}\}$.

{\bf Scattering amplitudes.} Scattering rate $\tau^{-1}_{mk}$ for 
 state with momentum $k$ in an $m$-subband
is related to corresponding self-energy $\Sigma_{mk}(\varepsilon)$:
$\tau^{-1}_{mk}=-2{\rm Im}\;\left\{\Sigma_{mk}\right\}$.
The current-carrying states are semiclassical, therefore 
within the Drude approximation the self-energies for these states are formally additive with respect to scattering at different impurities, also they  depend on $k$ only through the total energy:
\begin{align}
\Sigma_{mk}=\sum_i\Sigma_{mk}^{(i)},\quad \Sigma_{mk}^{(i)}\equiv\Sigma^{(i)}\left(E=\varepsilon_m+k^2/2m^*\right).
\nonumber
\end{align}
The self-energies can be expressed in terms of diagonal matrix elements of  renormalized scattering operator: $\Sigma^{(i)}_{m}=\tilde{V}^{(i){\rm (ren)}}_{m,m}.$
To evaluate $\tilde{V}^{(i){\rm (ren)}}_{m,m}$ we should single out transitions involving states within the $N$-band and take them into account nonperturbatively, using a fully quantum multi-impurity approach. The transitions between states with $m\neq N$ can be treated perturbatively and semiclassically. Let us introduce a composite perturbative amplitude for transition between two $m\neq N$ 
 states $|m_1\rangle$ and $|m_2\rangle$ due to scattering at an impurity $i$:
\begin{align}
\tilde{V}^{(i)}_{m_1,m_2}=V^{(i)}_{m_1,m_2}+V^{(i)}_{m_1,N}G
_{\varepsilon}(z_i,z_i)V^{(i)} _{N,m_2}\equiv\nonumber\\\equiv\frac{\tilde{\lambda}_i}{\pi^2}e^{i\phi_i(m_1-m_2)},
\quad\tilde{\lambda}_i=\lambda\left\{1+\frac{\lambda}{\pi^2 }G
_{\varepsilon}(z_i,z_i)\right\}.\label{nonres_scat_tubepoll1}
\end{align}
  The first term on the right hand side of \eqref{nonres_scat_tubepoll1} describes the direct transitions between two $m\neq N$ states, while the second term describes composite scattering processes with excursions to the   $N$-subband. The latter excursions are treated nonperturbatively in terms of the exact  Green function $G_{\varepsilon}(z,z')$ for the purely one-dimensional motion of an electron in the field of impurities.

 The semiclassical approximation fails for the states of the $N$-band in the immediate vicinity of the Van Hove singularity, where $|\varepsilon|\lesssim \varepsilon_{\min}^{\rm (nB)}$. These states enter the composite renormalized scattering amplitudes as intermediate states. When in the $N$-band, an electron has very small longitudinal momentum $k\sim\sqrt{\varepsilon}$ and, if $kn^{-1}<1$, it may simultaneously feel many impurities.  As a consequence, the Green function $G _{\varepsilon}(z_i,z_i)$ and, therefore, $\tilde{V}^{(i){\rm (ren)}}_{m_1,m_2}$ may depend  on the relative positions of  impurities $z_i-z_j$, strictly speaking, for all $j$. As we will see,  however, the set of relevant additional impurities is reduced to a pair of the closest ones, $j=i\pm 1$, so that
$\tilde{\lambda}^{(i)}=\tilde{\lambda}^{(i)}\left(L^{(+)}_i,L^{(-)}_i\right)$, where $L^{(\pm)}_i=|z_{i\pm1}-z_i|$.

   To take into account multiple scattering processes and find the renormalized amplitudes, we should solve the Dyson equation
\begin{align}
\frac{\tilde{\Lambda}^{(i){\rm (ren)}}}{\pi^2}=\frac{\tilde{\lambda}^{(i)}}{\pi^2}+\frac{\tilde{\lambda}^{(i)}}{\pi^2}g_{\varepsilon}(0)\frac{\tilde{\Lambda}^{(i){\rm (ren)}}}{\pi^2},\\
g_{\varepsilon}(0)=\sum_{m\neq N}g_{\varepsilon}^{(m)}(0),\quad g_{\varepsilon}^{(m)}(0)=-\pi i\varepsilon^{-1/2}_m,
\label{nonres_scat_tube2}
\end{align}
where $g_{\varepsilon}^{(m)}(0)$
is the free one-dimensional Green function in the $m$-th subband. The summation in \eqref{nonres_scat_tube2} runs over $m\ne N$ because all excursions to the $N$-subband are already taken into account by the second term in \eqref{nonres_scat_tubepoll1}.
Having in mind the absence of resonant term in the sum \eqref{nonres_scat_tube2}, in the multi-channel case ($N\gg 1$) we can relate $g_{\varepsilon}({\bf r})$ to the Green function of a free two-dimensional electron and obtain
\begin{align}
g_{\varepsilon}(0)\approx -i\pi^2,\quad \tilde{\Lambda}^{{\rm (ren)}}_i=\tilde{\lambda}_i(1+i\tilde{\lambda}_i)^{-1}.
\label{nonres_scat_tube5}
\end{align}

The scattering rate can be directly expressed through the renormalized coupling constant:
\begin{align}
\frac{1}{\tau(\varepsilon)}=-\frac{2}{\pi^2}\sum_i{\rm Im}
\,\tilde{\Lambda}^{{\rm (ren)}}_i
=-\frac{2n}{\pi^2}\left\langle{\rm Im}
\,\tilde{\Lambda}^{{\rm (ren)}}_i
\right\rangle_{i},
\label{selfe1rew}
\end{align}
where  the coupling constant $\tilde{\Lambda}^{{\rm (ren)}}_i$ depends on $i$ through the dependence  of $\tilde{\lambda}^{(i)}$ on $L^{(\pm)}_i$.

Now, substituting \eqref{nonres_scat_tubepoll1} to \eqref{nonres_scat_tube5} and using $\lambda\ll 1$, we finally obtain  the renormalized scattering amplitude
\begin{align}
\tilde{\Lambda}^{{\rm (ren)}}_i=\lambda(q_i^{-1}+1+i\lambda)^{-1},\label{nonres_scat_tube5as}\\ q_i=-\left[(\lambda/\pi^2) G_{\varepsilon}(z_i,z_i)\right]^{-1}-1.
\label{nonres_scat_tube5ask}
\end{align}
We see that the result for scattering amplitude $\tilde{\Lambda}^{{\rm (ren)}}_i$ (and, therefore, for resistivity) is expressed in terms of a {\it single-particle} Green function of a strictly one-dimensional problem, while for resistivity of a real one-dimensional system one would need a {\it two-particle} Green function. This simplification occurs in our multi-channel case due to the fact that the states in the resonant band do not contribute to the current directly.

{\bf Exact one-dimensional Green function.} The function $G_{\varepsilon}
$ is governed by the Schr\"{o}dinger equation:
\begin{eqnarray}
\left\{-\frac{1}{(2\pi)^2}\frac{d^2}{dz^2}+U(z)-\varepsilon\right\}G(z,z_i)=-\delta(z-z_i),\label{Schrodinger_eq}\\
U(z)=\lambda/\pi^2\sum_j\delta(z-z_j).
\label{Schrodinger_eq1}
\end{eqnarray}
Strictly speaking, one has to solve \eqref{Schrodinger_eq} for arbitrary number of impurities. As we have shown in \cite{IosPeshPRB2020}, under semiclassical conditions, i.e. for $\varepsilon\gg \varepsilon_{\min}$, the single-impurity approach is justified: one has to keep only the term $j=i$ in the sum. For $\varepsilon< \varepsilon_{\min}$, the multi-impurity effects become crucial. However, as we show below, these effects are mostly reduced to just two-impurity ones! It is enough to consider the nearest neighbor impurities, so that we have to substitute
\begin{align}
\overline{U}(z)=\lambda/\pi^2\sum_{j=i,i\pm1}\delta(z-z_j)
\label{Schrodinger_eq_simplified}
\end{align}
instead of $U(z)$ in \eqref{Schrodinger_eq}.
Other impurities (i.e., those with $j=i\pm2,i\pm3,\ldots$) play only a secondary role (see below).

The solution of equation \eqref{Schrodinger_eq} with the truncated potential \eqref{Schrodinger_eq_simplified}   leads (after substitution to \eqref{nonres_scat_tube5ask}) to an additive result: $q_i=q_i^{(+)}+q_i^{(-)}$, where
\begin{align}
q_i^{(\pm)}(\mbox{two impurities})\approx(k/4\lambda)\cot k\left[L^{(\pm)}_i+1/4\lambda\right]+\nonumber\\+i\left(k/4\lambda\right)^3\cot^2 k\left[L^{(\pm)}_i+1/4\lambda\right],\qquad k=2\pi\sqrt{\varepsilon}.\nonumber
\end{align}
It can be shown that taking into account distant impurities leads to only small correction to ${\rm Re}\,q_i$, $\delta {\rm Re}\,q_i\sim (k/4\lambda)^{2}$.
At the same time they ensure a dramatic suppression of ${\rm Im}\,q_i$. 
Indeed, ${\rm Im}\,q_i$ is responsible for the decay of the localized one-dimensional states due to  tunneling through a sequence of strongly repulsing impurities. Each impurity constitutes an additional potential barrier for tunneling electron and, therefore, suppresses the decay. In the limit of an infinite chain of impurities all the states are localized and ${\rm Im}\,q_i=0$.

We conclude that, due to localized character of states of the  
hamiltonian \eqref{Schrodinger_eq}, the exact $q_i$ is purely real, and
\begin{align}
q_i^{(\pm)}=q(L^{(\pm)}_i)\approx(k/4\lambda)\cot k\left[L^{(\pm)}_i+1/4\lambda\right].
 \label{ham07arc}
\end{align}
For $\varepsilon<0$ one should replace $k\to\kappa=2\pi\sqrt{|\varepsilon|}$, $\cot\to\coth$.

{\bf The resistivity: general result.} The averaging  
in \eqref{selfe1rew} is reduced to averaging over $L_i^{(\pm)}$:
\begin{align}
\frac{\rho}{\rho_0}=-\frac{1}{\lambda^2}{\rm Im}\,\langle\Lambda^{\rm (ren)}\rangle_{L^{(\pm)}}
= \nonumber\\=
\int_{0}^{\infty}\frac{\exp\{-n(L^{(+)}+L^{(-)})\}n^2dL^{(+)}dL^{(-)}}{([q(L^{(+)})+q(L^{(-)})]^{-1}+1)^2+\lambda^2}.
\label{genn1}
\end{align}
In principle, \eqref{genn1} together with \eqref{ham07arc} solve our problem: what is left is only to perform a double integration in \eqref{genn1}. Below we do it in different energy domains.

{\bf Resonant scattering.} For positive (and not very small) $\varepsilon$ the principal contribution to the resistivity comes from pairs of impurities with certain resonant distances, corresponding to poles of the scattering amplitude. 
It is highly improbable to have both $L_i^{(+)}$ and $L_i^{(-)}$ at resonance, so we can take into account only one of the two contributions $q_i^{(\pm)}$ and expand the cotangent near some point $kL=\pi p$, where $p=1,2,\ldots$ and $L$ is either $L_i^{(+)}$ or $L_i^{(-)}$. As a result, for resonant pairs  with $L$ close to $L_p(\varepsilon)$ we obtain
\begin{align}
\tilde{\Lambda}^{{\rm (ren)}}_i(L)\approx\left[4(L-L_p(\varepsilon))+i\right]^{-1},\label{ham07art13}\\
L_p(\varepsilon)=p/2\sqrt{\varepsilon}-1/2\lambda.\label{ham07art14}
\end{align}
The contribution of these poles
\begin{align}
\frac{\rho_{\rm res}}{\rho_0}=\frac{\pi n}{2\lambda^2}\sum_{p=1}^{\infty}e^{-nL_p}
=\frac{\pi n}{2\lambda^2}\left[\exp\left(\frac{n}{2\sqrt{\varepsilon}}\right)-1\right]^{-1}.
\label{serw1oon}
\end{align}
In terms of  $u,u_0$ this result takes the form \eqref{resuu01}.

{\bf Nonresonant scattering.}
There are several nonresonant contributions to the resistivity. For all of them the $\lambda^2$ term in the denominator of \eqref{genn1} can be neglected. The most important contribution $\rho_{\rm twin}$ comes from small $L\sim 1/\lambda\ll 1/n$. Taking into account only the leading  term in the small $L$ expansion of $q(L)$ we easily get from \eqref{genn1}:
\begin{align}
\frac{\rho_{\rm twin}}{\rho_0}
\approx2n\int_0^{\infty}\frac{e^{-nL}dL}{(4\lambda L+2)^2}=\frac{n}{4\lambda}.
\label{serw1oon1}
\end{align}
It is a result of scattering on ``twins'' -- rare pairs of impurities with non-typically small spatial separation $\sim 1/\lambda$  between them. The corresponding contribution to $\rho$ exists for both signs of $\varepsilon$ and does not depend on $\varepsilon$.

 The contribution $\rho_{\rm typ}$ of typical pairs with $L\sim1/n$  dominates only at large negative energies $u_0^{1/2}\ll u\ll u_0$ where the single-impurity approximation is applicable.

The logarithmic term $\Delta\rho_{\rm log}$, arising from pairs with $L$ in the interval $(1/\lambda,\min\{1/n, 1/\kappa\})$, is small compared to $\rho_{\rm twin}$, but, in contrast with $\rho_{\rm twin}$, it is $\varepsilon$-dependent.  An explicit evaluation of both $\rho_{\rm typ}$ and $\Delta\rho_{\rm log}$ is presented in  \cite{supplement}.

{\bf Crossover energies.}
The contribution of twins \eqref{twinies} coexists with the contribution of typical pairs \eqref{resuu02} in the range $\varepsilon<0$, $|\varepsilon|\gg\varepsilon_{\min}$ and with the resonant contribution \eqref{resuu01} in the range $\varepsilon>0$.
 Where the resistivity is dominated by twins scattering? For $\varepsilon>0$ the crossover between $\rho_{\rm res}$ and $\rho_{\rm twin}$ takes place at $\varepsilon\ll\varepsilon_{\min}$ so we can use the lower line asymptotics of \eqref{resuu01} and get
\begin{align}
\varepsilon_c^{(+)}\approx\varepsilon_{\min}\ln^{-2}(1/\lambda)\ll \varepsilon_{\min}.
\end{align}
For $\varepsilon<0$ the crossover occurs at $|\varepsilon|\gg\varepsilon_{\min}$, so that
\begin{align}
\varepsilon_c^{(-)}\approx -(1/4)(\varepsilon_{\min}\varepsilon_{\rm nB})^{1/2},\quad |\varepsilon_c^{(-)}|\gg\varepsilon_{\min}.
\end{align}
Note that the domain \eqref{interval} where the twins scattering dominates, spans mostly in the  region $\varepsilon<0$.
The minimum of resistivity lies at small positive energy, close to $\varepsilon_c^{(+)}$.

{\bf Discussion.} Why the scattering at twin impurities is dominant at low energy? In short, the answer is as follows: There is no special enhancement for the twin impurities scattering  at low energies, but there is a special suppression of single-impurity scattering: 
the non-Born effects lead to screening of the scattering amplitude $\Lambda_i^{\rm (ren)}$ at $\varepsilon\to 0$. This screening effect is, however, gradually destroyed, as a pair of impurities approach each other: the closer impurities, the weaker the screening. As a result, at smallest $\varepsilon$ scattering is dominated by twins -- anomalously close pairs of impurities. Below we discuss this issue in more detail.

As we have seen, 
 $\rho_{\rm twin}$ does not depend on $\varepsilon$, so  we may assume  
 $\varepsilon=0$ and get $q_i^{(\pm)}=[4\lambda L_i^{(\pm)}+1]^{-1}$ and
\begin{align}
\rho_{\rm twin}(L)/\rho_0\approx
\left(4\lambda L+2\right)^{-2}
\label{ham07cuomo97}
\end{align}
Although the result \eqref{ham07cuomo97} formally explains the dominance of twins scattering, 
it is instructive to discuss the following paradox:  at first glance  a twin pair should behave as a single composite impurity with a cumulated coupling constant $\lambda_{\rm twin}=2\lambda$ and, being a point-like object, should experience the same type of screening, as a solitary impurity. Let us argue that this assertion is only partly correct.

As long as 
only transitions within the $N$-band is considered, the phases of scattering amplitudes at neighbouring impurities $(i,j)$ do not differ  
 and a twin pair indeed can be treated as a single composite impurity. 
If it were true also for the interband processes then the entire process of twins scattering could be reduced to an effective single-impurity one and the suppression of the single-impurity scattering  
would also apply to twins.
However, when one considers the transitions between states from $m,m^{\prime}\neq N$ bands  
then the matrix elements for scatterings at different components of the pair  have different phases and the corresponding interference cross-terms rapidly oscillate:  
\begin{align}
|V_{mm^{\prime}}^{(ij)}|^2=|V_{mm^{\prime}}^{(i)}|^2+|V_{mm^{\prime}}^{(j)}|^2+2{\rm Re}\,\left[V_{mm^{\prime}}^{(j)}V_{mm^{\prime}}^{(i)*}\right],\nonumber\\
V_{mm^{\prime}}^{(j)}V_{mm^{\prime}}^{(i)*}\propto\exp\{ i(m-m^{\prime})\Delta\phi+ik_{m m^{\prime}}L\}
,\nonumber
\end{align}
where $L=|z_i-z_j|$, $\Delta\phi=\phi_i-\phi_j$ and a typical  momentum transfer $k_{m m^{\prime}}\sim N\gg 1$. Oscillations lead to suppression of the cross-terms after averaging over $L$, $\Delta\phi$.

Suppression of the interference terms destroys the full equivalence of twin pairs to solitary composite impurities and, therefore, makes the statement about the screening of the scattering at twins invalid. 
Thus,  the dominant role of twins scattering is an essentially {\it quasi}-one-dimensional (multi-channel, $N\gg 1$) effect, it  can not be observed in a purely one-dimensional system, where $N=1$.

This work was supported by Basic Research Program of The Higher School of Economics. The research of N. Peshcherenko was also supported by the Foundation for Advancement of Theoretical Physics and Mathematics ``Basis'' under grant 20-1-5-150-1. The authors are indebted to I. S. Burmistrov and P. M. Ostrovsky for valuable comments.

\begin{widetext}
\begin{center}
{\large \bf Two-impurity scattering  in quasi-one-dimensional systems: \\Supplemental material}
\end{center}
\setcounter{equation}{0}
\renewcommand{\theequation}{S\arabic{equation}}
\end{widetext}

{\bf Evaluation of nonresonant scattering contributions $\rho_{\rm typ}$ and $\Delta\rho_{\rm log}$.}
As we have seen, for evaluation of $\rho_{\rm twin}$ it is enough to expand $\cot$ (for $\varepsilon>0$) or $\coth$ (for $\varepsilon<0$) to the lowest order in its argument, but, on the other hand, the $1/4\lambda$ shift of this argument, as well as the unity term in the denominator, are essential. For finding the rest of nonresonant contributions, (i.e., for $\rho_{\rm typ}$ and $\Delta\rho_{\rm log}$), the expansion of $\coth$ is not sufficient, so one should keep  $\coth$ as it is. However, there are some alternative simplifications that allow for the evaluation of these terms. Namely, one can  neglect unity in the denominator of  (22) and the shift $1/4\lambda$ of the arguments of $\cot$. The reminiscence of the latter shift, however, appears as an ultraviolet cutoff of certain logarithmically divergent integrals. We have to note that thus proposed simplified method of calculation does not allow for reliable determination of the numerical factors under logarithms, more sophisticated approaches are needed for that end. Anyway, the contributions of these numerical factors are relatively small.

Having in mind that the above approximation does not work for $\rho_{\rm twin}$, we should first eliminate the corresponding part from the integrand. Then for $\Delta\rho\equiv\rho-\rho_{\rm twin}$ we get
\begin{widetext}
\begin{align}
\frac{\Delta\rho}{\rho_0}\approx
\int_{0}^{\infty}e^{-n(L+L^{\prime})}n^2dLdL^{\prime}[(q(L)+q(L^{\prime})]^2\approx
2\int_{0}^{\infty}e^{-nL}\left[q^2(L)-(4\lambda L)^{-2}\right]ndL+
2\left[\int_{1/2\lambda}^{\infty}e^{-nL}q(L)ndL\right]^2
\approx\nonumber\\\approx\frac{1}{8u_0^2}\left[2uF_2(u)+\left[F_1(u)+\ln u_0\right]^2\right],\qquad F_m(u)=\int_{0}^{\infty}e^{-t/2u}\left(\coth^m t-\frac{1}{t^m}\right)dt,\qquad u=\frac{\pi\sqrt{-\varepsilon}}{n}.
\nonumber
\end{align}
The functions $F_{1,2}(u)$ can be expressed in terms of digamma function $\psi$:
\begin{align}
F_1(u)=\ln(1/4u)-\psi\left(1/4u\right)-2u,\nonumber\qquad
F_2(u)=1+2u+(1/2u)\ln 4u+(1/2u)\psi\left(1/4u\right).
\end{align}
Their asymptotics can be easily found
\begin{align}
F_2(u)\approx\left\{\begin{aligned}2u-1+(1/2u)\ln u,\quad & u\gg1,\\
4u/3,\quad & u\ll1,
\end{aligned}
\right.\quad F_1(u)\approx\left\{\begin{aligned}2u-\ln u,\quad & u\gg1,\\
(4/3)u^2,\quad & u\ll1.
\end{aligned}
\right.\nonumber
\end{align}
We should stress that $u\ll 1$ asymptotics are valid for both signs of $\varepsilon$, while the obtained $u\gg 1$  asymptotics are applicable only for $\varepsilon<0$ case. The case $\varepsilon>0$ and $u\gg 1$ is not relevant, since the nonresonant scattering is marginal in this energy domain.

To single out logarithmic contributions in $F_1(u)$, $uF_2(u)$ (which are related to $\Delta\rho_{\rm log}$) it is convenient to introduce the following modified functions:
\begin{align}
 \tilde{F}_1(u)=\frac14\ln(1+u^4)+\int_{0}^{\infty}e^{-t/2u}\left(\coth t-\frac{1}{t}\right)dt\approx\left\{\begin{aligned}2u,\quad & u\gg1,\\
(4/3)u^2,\quad & u\ll1,
\end{aligned}
\right.
\label{choice}
\end{align}
\begin{align}
\tilde{F}_2(u)=-\frac{1}{8(1+u^4)^{1/4}}\ln(1+u^4)+\int_{0}^{\infty}e^{-t/2u}\left(\coth^2 t-\frac{1}{t^2}\right)dt\approx\left\{\begin{aligned}2u,\quad & u\gg1,\\
4u/3,\quad & u\ll1,
\end{aligned}
\right.
\label{choice2}
\end{align}
which, in contrast with $F_1(u)$, $uF_2(u)$, are logarithm-free.  Now the logarithmic contributions that belong to $\Delta\rho_{\rm log}$ can be easily separated from the logarithm-free ones, belonging to $\rho_{\rm typ}$:
\begin{align}
\Delta\rho=\rho_{\rm typ}+\Delta\rho_{\rm log},\quad \frac{\Delta\rho_{\rm typ}}{\rho_0}\approx\frac{1}{8u_0^2}\left[2u\tilde{F}_2(u)+\tilde{F}_1^2(u)\right]\approx\frac{u^2}{u_0^2}\left\{\begin{aligned}1,\quad & u\gg1,\\
8/3,\quad & u\ll1,
\end{aligned}
\right.\\
\frac{\Delta\rho_{\rm log}}{\rho_0}\approx\frac{1}{8u_0^2}\left[\frac{u}{4(1+u^4)^{1/4}}\ln(1+u^4)+2\tilde{F}_1(u)\ln\left[\frac{u_0}{(1+u^4)^{1/4}}\right]+\ln^2\left[\frac{u_0}{(1+u^4)^{1/4}}\right]\right]\approx\nonumber\\\approx\frac{1}{u_0^2}\left\{\begin{aligned}(u/2)\ln\left(u_0/u\right)+\ln^2\left(u_0/u\right),\quad & u\gg1,\\
(u^2/3)\ln u_0+(1/8)\ln^2 u_0,\quad & u\ll1.
\end{aligned}
\right.
\end{align}
Note that the choice \eqref{choice}, \eqref{choice2} is not unique: in principle, one could have introduced the ``counter-logarithm'' terms in a form $\frac{1}{\alpha}\ln(1+u^\alpha)$ for $\tilde{F}_1(u)$ and $\frac{1}{2\alpha\left(1+u^{\alpha}\right)^{1/\alpha}}\ln(1+u^\alpha)$ for $\tilde{F}_2(u)$ with any $\alpha>1$. We have chosen rather high $\alpha=4$ in order not to affect the $u\ll 1$ asymptotics.
\end{widetext}


\begin{thebibliography}{99}

\bibitem{VanHove} P.~Y.~Yu and M.~Cardona {\it Fundamentals of Semiconductors.  Physics and Material Properties.} Chap. 6.2, Springer, (2010).
\bibitem{single-wall0}  Z. Zhang, D. A. Dikin, R. S. Ruoff, and V. Chandrasekhar, Europhysics Letters, {\bf 68}, 713 (2004).
\bibitem{single-wall}
B. Babi\'{c} and C. Sch\"{o}nenberger, Phys. Rev. {\bf B 70}, 195408 (2004)
\bibitem{multi-wall1}
 J. Kim, J. R. Kim, Jeong-O Lee, J. W. Park, H. M. So, N. Kim, K. Kang, K. H. Yoo, and J. J. Kim, Phys. Rev. Lett. {\bf 90}, 166403 (2003).
\bibitem{multi-wall2} W. Yi, L. Lu, H. Hu, Z. W. Pan, and S. S. Xie, Phys. Rev. Lett. {\bf 91}, 076801 (2003).
\bibitem{Brandt77}   N.~B.~Brandt, D.~V.~Gitsu, A.~A.~Nikolaeva, and Ya.~G.~Ponomarev,  
Sov. Phys. JETP, {\bf 45}, 1226 (1977).
\bibitem{Nikolaeva2008}  A.~Nikolaeva, D.~Gitsu, L.~Konopko, M.~J.~Graf, T.~E.~Huber, Phys. Rev. {\bf B 77}, 075332 (2008)
\bibitem{nanoribbon1} K.~Wakabayashi, Phys. Rev. {\bf B 64}, 125428 (2001)
\bibitem{nanoribbon2} 
M.~M.~Pour, A.~Lashkov, A.~Radocea, X.~Liu, T.~Sun, A.~Lipatov, R.~A.~Korlacki, M.~Shekhirev, N.~R.~Aluru, J.~W.~Lyding, V.~Sysoev, A.~Sinitskii,
 Nature Communications {\bf 8}, 820, (2017).
\bibitem{constrictions} G.~Bastard, J.~A.~Brum, R.~Ferreira, Electronic States in
Semiconductor Heterostructures, in: Solid State Physics, Vol. {\bf 44}, 229 (1991), Ch. 8.
\bibitem{Thornton1986}
T.~J.~Thornton, M.~Pepper, H.~Ahmed,
D.~Andrews, G.~J.~Davies, Phys. Rev. Lett., {\bf 56}, 1198 (1986)
\bibitem{Zheng1986} H.~Z.~Zheng, H.~P.~Wei,  D.~C.~Tsui,
G.~Weimann, Phys. Rev. {\bf B 34}, 5635, (1986)
\bibitem{stripstudyexper} J.~C.~Chen, Yiping~Lin, Kuan Ting Lin, T.~Ueda, and S.~Komiyama, Applied Physics Letters {\bf 94}, 012105 (2009)
\bibitem{fano} U. Fano, Phys. Rev. {\bf 124}, 1866 (1961).
\bibitem{fano1}  A. E. Miroshnichenko, S. Flach, Y.~S.~Kivshar, Rev. Mod. Phys.  {\bf 82}, 2257 (2010)
\bibitem{IosPeshJETPL2018} A.~S.~Ioselevich and N.~S.~Peshcherenko, JETP Letters, {\bf 108}, 12, 825 (2018).


\bibitem{IosPeshPRB2019} A.~S.~Ioselevich and N.~S.~Peshcherenko, Phys. Rev. {\bf B 99},   035414 (2019).
\bibitem{IosPeshPRB2020} N.~S.~Peshcherenko and  A.~S.~Ioselevich,  Phys. Rev. {\bf B 102}, 134208 (2020).




\bibitem{LifshitsGredeskulPastur1982}
I. M. Lifshits, S. A. Gredeskul, L. A. Pastur
{\it Introduction to the Theory of Disordered Systems }, Wiley-VCH (1988).
\bibitem{BychkovDykhne1966}
Yu.~A.~Bychkov, A.~M.~Dykhne, 
JETP Letters, {\bf 3}, 202 (1966).
\bibitem{BychkovDykhne1967}
Yu.~A.~Bychkov, A.~M.~Dykhne,
JETP, {\bf 24}, 1285 (1967).
\bibitem{Ado2016}  I. A. Ado, I. A. Dmitriev, P. M. Ostrovsky, and M. Titov
Phys. Rev. Lett. {\bf 117}, 046601 (2016)
\bibitem{Ado2020} I. A. Ado, I. A. Dmitriev, P. M. Ostrovsky, and M. Titov
Phys. Rev. Lett. {\bf 124}, 259902 (2020).
\bibitem{KumBag1991} A.~Kumar and P.~F.~Bagwell, Phys. Rev. {\bf B 43},   9012 (1991).
\bibitem{supplement} See Supplemental Material for technical details.

\end{thebibliography}
\end{document}